\documentclass[%
preprint,
 amsmath,amssymb,
 aip,
]{revtex4-2}
\usepackage[dvipsnames]{xcolor}
\usepackage{graphicx}% Include figure files
\usepackage{dcolumn}% Align table columns on decimal point
\usepackage{bm}% bold math
\begin{document}

\preprint{APS/123-QED}

\title{Collective dynamics in a one-dimensional Heisenberg ferromagnetic spin chain}% Force line breaks with \\

\author{R. Arun}%
 \email{arunbdu@gmail.com}
\affiliation{ Center for Nonlinear and Complex Networks, SRM TRP Engineering College, Tiruchirappalli–621 105, Tamil Nadu, India.}%
\affiliation{  Centre for Research, Trichy SRM Medical College Hospital and Research Center, Tiruchirappalli 621 105, Tamil Nadu, India.}%
\author{M. Lakshmanan}
\email{lakshman.cnld@gmail.com}
\affiliation{%
	Department of Nonlinear Dynamics, Bharathidasan University, Tiruchirappalli - 620024, India.
}%

\author{Avadh Saxena}
 \email{avadh@lanl.gov}
\affiliation{
Theoretical Division and Center for Nonlinear Studies, Los Alamos National Laboratory, NM 87545, USA
}%
%\affiliation{%

\date{\today}% It is always \today, today,

\begin{abstract}
We investigate the different oscillatory modes, namely, complete synchronization, inphase synchronization, antiphase synchronization and desynchronization in a one-dimensional anisotropic Heisenberg ferromagnetic spin chain consisting of a large number of spins. By solving the associated Landau-Lifshitz-Gilbert-Slonczewski equation for the spins we show the simultaneous existence of the above mentioned oscillatory modes in the spins.  We observe that when the number of the spins is large the synchronization is lost between the spins; however, we identify that the field-like torque is able to induce synchronous oscillations of the spins in the chain again. We also confirm the agreement of the numerically obtained values of the frequency of the inphase synchronized oscillations with the analytically obtained values.  %\begin{description}
%\end{description}
\end{abstract}

\maketitle
%\tableofcontents
\section{Introduction}
Understanding the intrinsic localized modes and associated oscillations of different types in the context of  classical nonlinear Hamiltonian spin lattices has been a fundamentally interesting topic. 
The investigation on the dynamics of classical Heisenberg ferromagnetic spin chain has always been a problem  of substantial interest. Anisotropic Heisenberg ferromagnetic spin system is one of the important discrete nonlinear dynamical systems which has received considerable attention in diverse areas of physics for a long time. Obtaining exact solutions for discrete nonlinear physical systems remains a significant and challenging problem of current research interest. In the continuum limit, the one-dimensional Heisenberg ferromagnetic spin system with nearest-neighbor exchange interaction is known to give rise to several completely integrable systems that support soliton solutions. For instance, the pure isotropic system \cite{laksh_1997,takh,zakh}, the uniaxial anisotropic system \cite{boro,naka} and the biaxial anisotropic system \cite{skly} all turn out to be completely integrable infinite dimensional nonlinear systems.

Apart from a variant called Ishimori spin chain there are no other exactly solvable discrete dynamical Heisenberg spin systems that have been identified in the literature. The Ishimori model \cite{ishi1} remains completely integrable despite its nonpolynomial structure with the Hamiltonian $\mathcal{H}$=-log$(1+\sum S_k \cdot S_{k+1})$. It possesses a Lax pair, an infinite hierarchy of conserved quantities, and admits exact analytical solutions through techniques such as the inverse scattering transform \cite{ishi}. In contrast, two of the present authors \cite{laksh_2008} have shown that the discrete lattice system, incorporating onsite anisotropy and an external magnetic field, admits several classes of exact solutions expressed through Jacobian elliptic functions \cite{laksh_2008}. The emergence of intrinsic localized breathers in appropriately anisotropic ferromagnetic spin chains is of considerable practical relevance \cite{sie,zolo}. Recently, Kamburova et al. have demonstrated that increasing either the strength of the exchange interaction or the magnetic anisotropy within a modified region of a spin chain gives rise to a potential barrier, leading to the transmission and reflection of an incident soliton \cite{kam}. In recent years, significant attention has been devoted to the identification and analysis of intrinsic localized modes (ILMs), or discrete breathers, in classical nonlinear Hamiltonian lattices, including magnetic chains \cite{sie2,page,camp,fla,mac,aub,rak,zolo2,sav,lai,eng,ngu}. ILMs are dynamical localized states, which include periodic oscillations in time in a localized space. Though the analytical solution for the discrete system with onsite anisotropy and an external magnetic field is available, a solution for the anisotropic ferromagnetic spin chain with external magnetic field along with damping is unavailable.  

Here, we delve into the different dynamical states present in the anisotropic Heisenberg ferromagnetic spin chain system. Starting from the Hamiltonian associated with the lattice corresponding to the Heisenberg spin chain we solve numerically the associated Landau-Lifshitz-Gilbert-Slonczewski (LLGS) equation, which includes the spin-transfer-torque effect on the spins associated with the lattice. We prove the existence of self oscillations, i.e. oscillations without any periodic input on the spins in the chain.   By numerically solving the LLGS equation obtained for the $N$ number of spins with nearest neighbour interaction we identify different dynamical states associated with the spin chain system.  Specifically, we identify different kinds of oscillations in the chain, namely  (i) completely synchronized oscillations, (ii) inphase synchronized oscillations, (iii) antiphase synchronized oscillations and (iv) desynchronized oscillations by changing the direction of external field in the presence of field-like torque. We observe the simultaneous existence of the above oscillatory modes in the chain consisting of $N$ number of spins. The synchonization between the different pairs of the spins is confirmed by the standard deviation and phase difference between them. The numerical results are validated with the frequency of the oscillations derived analytically. We confirm that when the field-like torque is incorporated in the chain, the spins are expected to show inphase synchronized oscillations from desynchronized oscillations.   

The plan of the paper is as follows: The model of the spin system is discussed in Sec. II, where the LLGS equation corresponding to the Heisenberg ferromagnetic spin chain is derived from the respective Hamiltonian of the lattice. In Sec. III, the existence of different modes of oscillations including complete synchronized, inphase synchronized, antiphase synchronized and desynchronized oscillations are investigated. The main results are summarized and concluded in Sec. IV.

\section{Model}
The Hamiltonian corresponding to the evolution of $N$ number of spins of a one-dimensional anisotropic Heisenberg ferromagnetic spin chain is given by
\begin{align}
	\mathcal{H} = &-\sum_{k=1}^N [A ~S_k^x S_{k+1}^x + B ~S_k^y S_{k+1} +C ~S_k^z S_{k+1}^z] - K_z\sum_{k=1}^N (S_k^z)^2-  {\bf H}\cdot \sum_{k=1}^N {\bf S}_k , \label{H1}
\end{align}
where $k$ runs from 1 to $N$. The edge spins at $k$ = 1 and $N$ are left open and there is no direct connection between them. In Eq. \eqref{H1}, $A,~B$ and $C$ are exchange interaction constants, while $K_z$ is an onsite anisotropy parameter along the $z$-direction. Further, ${\bf H}=H_x {\bf e}_x + H_y {\bf e}_y + H_z {\bf e}_z$ is the  external field. Here, ${\bf e}_x$, ${\bf e}_y$, and ${\bf e}_z$ are the unit vectors along the positive $x$, $y$, and $z$ directions, respectively. 
%Eq. \eqref{H1} can be rewritten as
%\begin{align}
%	\mathcal{H} = &-\sum_{i=1}^N[A~ S_i^x S_{i+1}^x + B ~S_i^y S_{i+1}^y + C ~S_i^z S_{i+1}^z]-K_z \sum_{i=1}^N(S_i^z)^2 - H_x \sum_{i=1}^N S_i^x- H_y \sum_{i=1}^N S_i^y- H_z \sum_{i=1}^N S_i^z . \label{H2}
%\end{align}
After expanding the Hamiltonian, given in Eq. \eqref{H1}, around the $k$-th spin we get
\begin{align}
	\mathcal{H} = &-[.....+A~S_{k-1}^x S_k^x + A~S_k^x S_{k+1}^x + B~S_{k-1}^y S_k^y+B~S_k^y S_{k+1}^y+C~S_{k-1}^z S_k^z + C~S_k^z S_{k+1}^z+.....]\nonumber\\
	&-K_z ~[.....+(S_{k-1}^z)^2+(S_k^z)^2+(S_{k+1}^z)^2+.....]-H_x[...+S_{k-1}^x+S_k^x+S_{k+1}^x+...]\nonumber\\
	&-H_y[...+S_{k-1}^y+S_k^y+S_{k+1}^y+...]-H_z[...+S_{k-1}^z+S_k^z+S_{k+1}^z+...].\label{H2}
\end{align}
The Hamiltonian given in Eq.\eqref{H2} is applicable for arbitrary boundary conditions.
The effective field for the $k$-th spin can be obtained from the Hamiltonian as
\begin{align}
	H_{eff,k}^x  = -\frac{\delta \mathcal H}{\delta S_k^x}= &A(S_{k-1}^x+S_{k+1}^x)+H_x,\label{eff1}\\
	H_{eff,k}^y = -\frac{\delta \mathcal H}{\delta S_k^y}= &B(S_{k-1}^y+S_{k+1}^y)+H_y,\label{eff2}\\
	H_{eff,k}^z= -\frac{\delta \mathcal H}{\delta S_k^z}= &C(S_{k-1}^z+S_{k+1}^z)+2K_z S_k^z+H_z.\label{eff3}
\end{align}
The effective fields for the first and $N$-th spins are given by
\begin{align}
H_{eff,1}^x &= AS_2^x+H_x, \hskip0.5cm H_{eff,N}^x = AS_{N-1}^x+H_x,\label{eff4}\\
H_{eff,1}^y &= BS_2^y+H_y, \hskip0.5cm  H_{eff,N}^y = AS_{N-1}^y+H_y,\label{eff5}\\
H_{eff,1}^z &= CS_{2}^z+2K_z S_1^z+H_z, \hskip0.5cm  H_{eff,N}^z = CS_{N-1}^z+2K_z S_N^z+H_z,\label{eff6}
\end{align}

By considering the one-dimensional spin chain, the dynamics of the $n$-th spin in the presence of current is governed by the following LLGS equation~\cite{laksh_2021,laksh_2018}
\begin{align}
	\frac{d{\bf S}_k}{dt} = -\gamma~ {\bf S}_k \times {\bf H}_{eff,k} - \alpha~ {\bf S}_k\times ({\bf S}_k\times {\bf H}_{eff,k})+ j ~{\bf S}_k\times({\bf S}_k\times {\bf e}_x) + j~\beta~ {\bf S}_k\times {\bf e}_x \label{llgs}.
\end{align}  

In Eq. \eqref{llgs} the effective field of the $k$-th spin ${\bf H}_{eff,k}$ is defined as ${\bf H}_{eff,k} = H_{eff,k}^x {\bf e}_x+ H_{eff,k}^y {\bf e}_y+ H_{eff,k}^z {\bf e}_z$.  Here, $\gamma$ is the gyromagnetic ratio, $\alpha$ is the damping constant, $j$ and $\beta$ are the strengths of spin-transfer torque and field-like torque, respectively.

%Henceforth, the motion of the spins are numerically investigated using Runge-Kutta-4 method with the following parameters $A$ = $B$ = $C$ = 1.0 and $K_x$ = $K_y$ = 0, $K_z$ = 1.0. These values will be maintained throughout the paper unless otherwise mentioned.

\section{Self-oscillations}
In the absence of anisotropy, Eq.\eqref{llgs} has analytically been solved for the single spin by the present authors \cite{laksh_2021}. However, it is hard to be solved analytically in the present form even for a single spin. Hence, in the following, the dynamics of the $N$  spins is numerically investigated using the Runge-Kutta-4 method and also by using Mathematica. 
Specifically, we have used the adaptive step-size method in Runge-Kutta-4 scheme, which is explained as follows: In each iteration the magnitude of the unit spin vector is checked and maintained with the condition $|S_k|^2 <(1+10^{-10})$ and $|S_k|^2 >(1-10^{-10})$. If the magnitude does not meet the conditions the step size is halved and the iteration is repeated again. Otherwise, the step size is doubled for the next iteration.  In this way, to maintain the accuracy of $|S_k|^2$ the step size is adjusted accordingly.
\subsection{Complete and inphase synchronized oscillations}
\begin{figure}[h!]
	\centering	\includegraphics[width=1\linewidth]{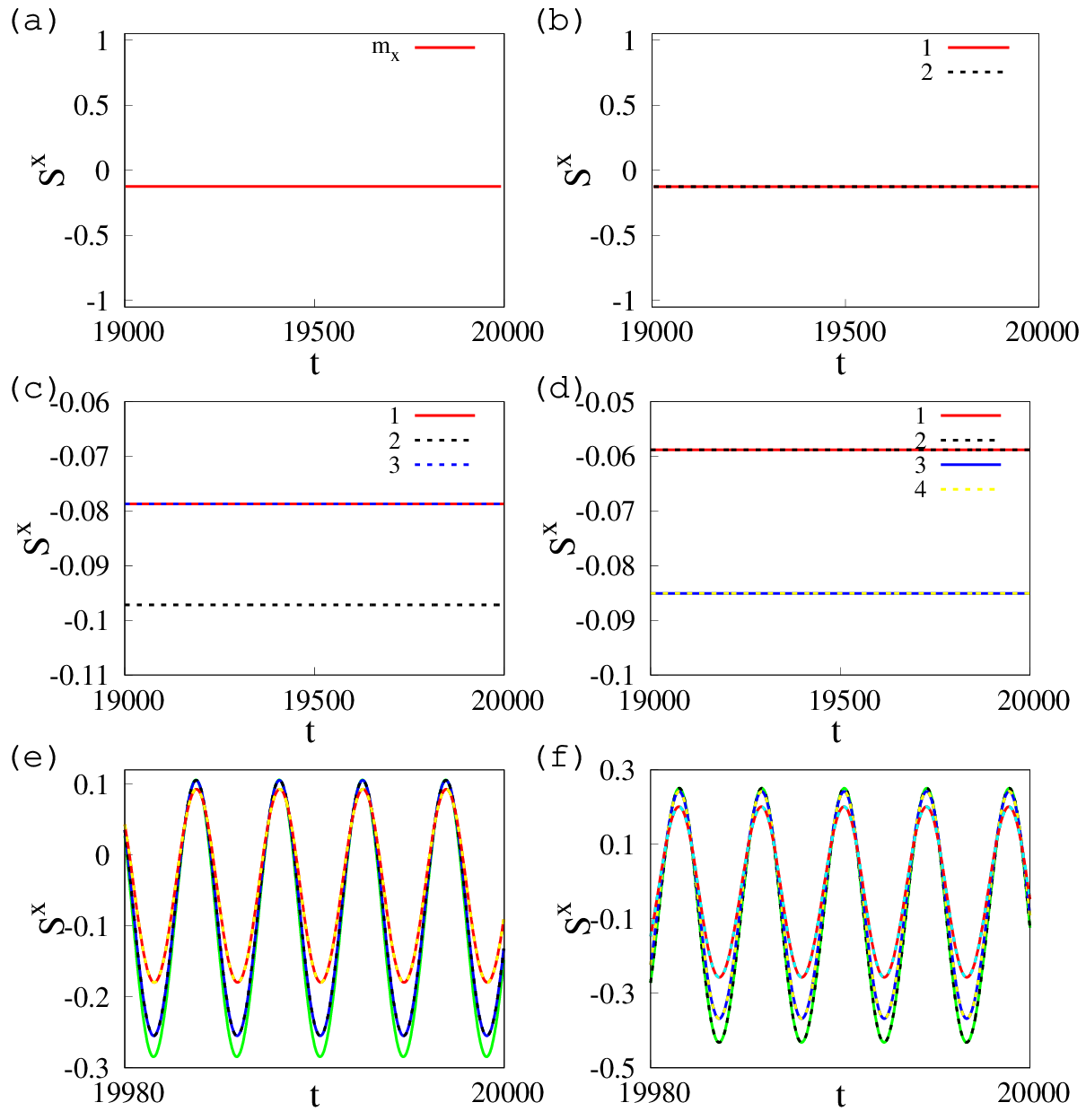}
	\caption{Time evolution of $S^x$ for the number of spins (a) 1, (b) 2, (c) 3, (d) 4, (e) 5, and (f) 6 for $\beta$ = 0. The color codes: (e) spins 1 (solid red), 2 (solid blue), 3 (solid green), 4 (dashed black), and 5 (dashed yellow).  (f) Spins 1 (solid red), 2 (solid blue), 3 (solid green), 4 (dashed black), 5 (dashed yellow), and 6 (dashed cyan).}
	\label{fig1}
\end{figure}

Here, we show the existence of self-oscillations, that is oscillations in the absence of any external periodic source, of the spins in the one-dimensional chain. The self-oscillations of the spins are demonstrated here for the values of the parameters $A=2,B=2,C=1,\gamma=1.0,K_z=0.1,H_x=0.1,H_y=0,H_z=0,\alpha=0.005,j=0.1,\beta=0$, and the condition ${\bf S}_p={\bf e}_x$, corresponding to the case where the fixed polarized vector ${\bf S}_p$ is parallel to the $x$-axis. In Fig. \ref{fig1}(a-f) we have plotted the evolution of $x$-components ($S^x$) for the number of spins 1, 2, ..., 6, respectively. From the figures we can observe that when the total number of spins is 4 and below ($N\leq 4$), no oscillations are exhibited and only steady states are possible. However, steady oscillations are exhibited while the total number of spins is above 4 ($N>4$) as shown in Figs. \ref{fig1}(e) and (f). From Figs. \ref{fig1}(e-f) we can notice that the pair of spins ($i,N+1-i$) exhibit synchronized oscillations. The amplitudes of the oscillations of spins 1, 2, and 3 are slightly different while their frequencies are the same. 

\begin{figure}[h!]
	\centering	\includegraphics[width=1\linewidth]{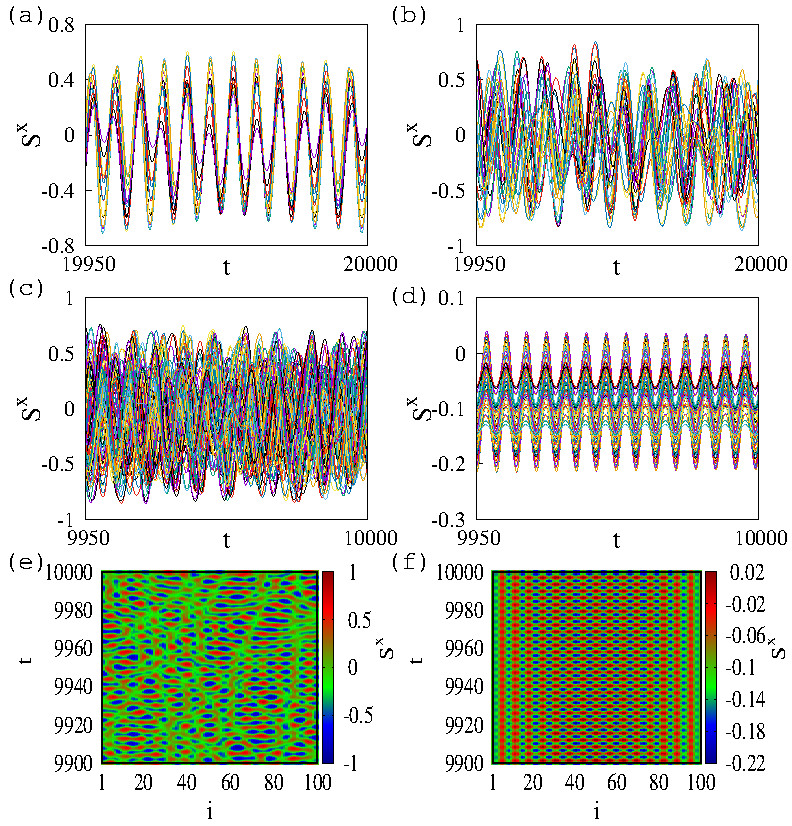}
	\caption{Time evolution of $S^x$ when $\beta$ = 0 for the number of spins $N$ (a) 25, (b) 30, (c) 100 and for (d) 100 when $\beta = -0.6$. Time evolution of 100 spins when (e) $\beta$ = 0 and (f) $\beta = -0.6$.}
	\label{fig2}
\end{figure}

\begin{figure}[h!]
	\centering	\includegraphics[width=0.8\linewidth]{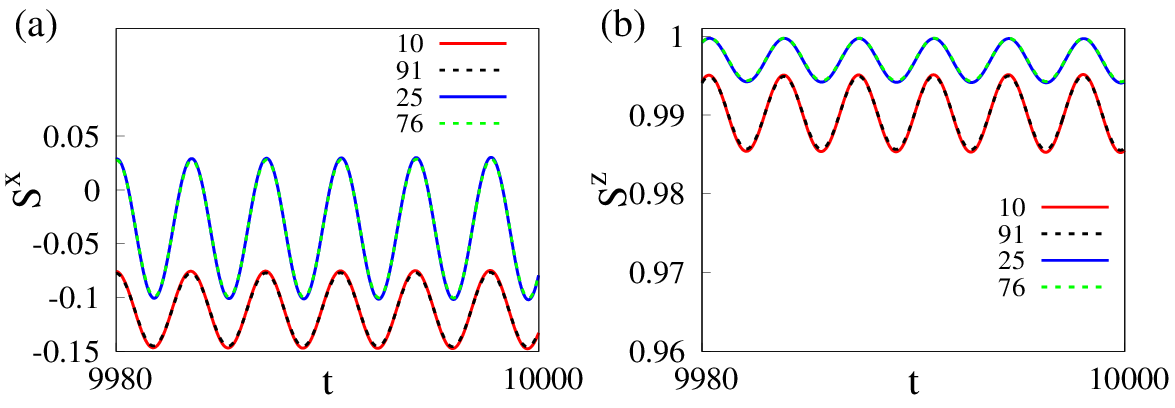}
	\caption{Complete synchronization between the spins (10,91) and (25,76) in their time evolution of (a) $S^x$, and (b) $S^z$ when $\beta = -0.6$.}
	\label{fig2A}
\end{figure}

To investigate the dynamics for a large number of spins, the time evolution of $S^x$ is plotted for 25, 30, and 100 number of spins in Figs. \ref{fig2}(a-c). As we can see from Figs. \ref{fig2}(a-c), the synchronization is lost in the oscillations when the number of spins is increased above 25 and the spins oscillate desynchronously. For the strength of the field-like torque $\beta = -0.6$, the spins exhibit inphase synchronized oscillations as shown in Fig. \ref{fig2}(d). This implies the possibility of getting synchronization due to the field-like torque. The frequency of oscillations of the 100 spins are the same and the amplitudes of the oscillations are different.  The time evolution of $S^x$ of 100 spins is plotted for $\beta$ = 0 and $\beta =-0.6$ in Figs. \ref{fig2}(e) and (f), respectively. Fig. \ref{fig2}(e) confirms the desynchronization in oscillations of the spins and in Fig. \ref{fig2}(f) we can observe the inphase synchronization between the spins due to the field-like torque. Also, we can verify the complete synchronization between ($i,N+1-i$) spins due to the field-like torque. This  complete synchronization is visualized in Fig. \ref{fig2A}, where the time evolution of the spins 10, 91, 25, and 76 are plotted for $S^x$ and $S^z$ in Fig. \ref{fig2A}(a) and (b), respectively, and we can notice that the oscillations of the spins 10 and 91 as well as 25 and 76 are completely synchronized.

\begin{figure}[h!]
	\centering	\includegraphics[width=0.9\linewidth]{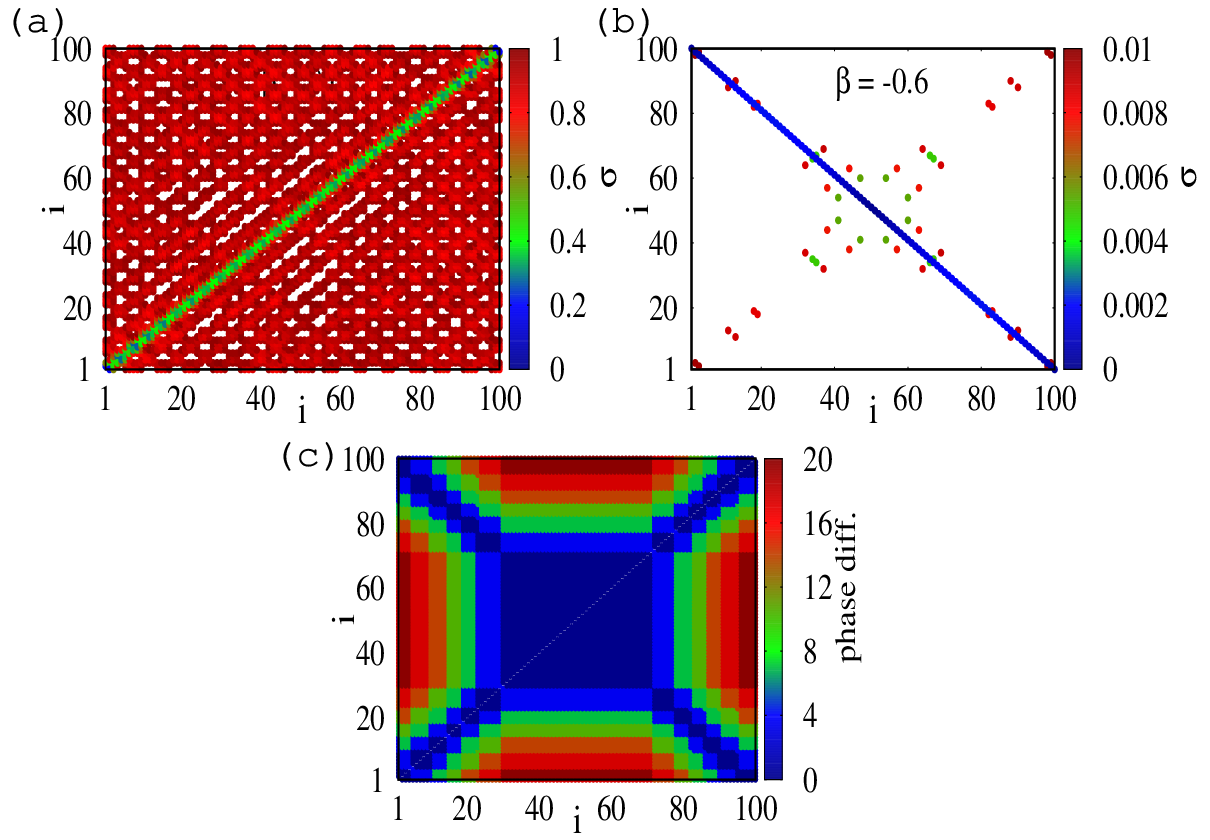}
	\caption{The standard deviation estimated for $S^x(t)$ between different pairs of spins `$i$' and `$N+1-i$' for (a) $\beta$ = 0 and (b) $\beta = -0.6$. (c) Phase difference in degrees between different the pairs of spins when $\beta = -0.6$. Here, $H_x=0.1,H_y=0, H_z=0, j=0.1$.}
	\label{fig3}
\end{figure}

To verify the complete synchronization between all the pair of spins (1,100), (2,99), (3,98), ... , (100,1) the standard deviation between their respective $S^x(t)$ is calculated and plotted in Figs. \ref{fig3}(a) and (b) for $\beta$ = 0 and $\beta = -0.6$, respectively.
The standard deviation between the $j$-th and $k$-th spins can be calculated during the time between $t=t_1$ and $t=t_2$, which is given as
\begin{align}
	\sigma = \frac{1}{t_2-t_1} \sum_i\sqrt{ \frac{[S^x_j(t_i)-\mu^x(t_i)]^2+[S^x_k(t_i)-\mu^x(t_i)]^2}{2}}
\end{align}
where $t_i$ = $t_1,~t_1+h,~t_1+2h,~...,t_2$. Here $h$ is the step size in the computation and $\mu^x(t_i) = (S^x_j(t_i)+S^x_k(t_i))/2$.

In Fig. \ref{fig3}(a) and (b), where the standard deviation $\sigma$ is plotted for $\beta$ = 0 and $-0.6$, respectively, the white regions in Figs. \ref{fig3}(a) and (b) correspond to $\sigma>1.0$ and $\sigma>0.01$, respectively.  In Figs. \ref{fig3}(a) we can observe that the standard deviation is quite high between any pair of spins in the chain for $\beta$ = 0. When $\beta = -0.6$, the standard deviation $\sigma$ between the spins ($i,N+1-i$) is close to zero as shown in Fig. \ref{fig3}(b). This implies that these pairs of spins are exhibiting completely synchronized oscillations due to the field-like torque.

\begin{figure}[h!]
	\centering	\includegraphics[width=0.6\linewidth]{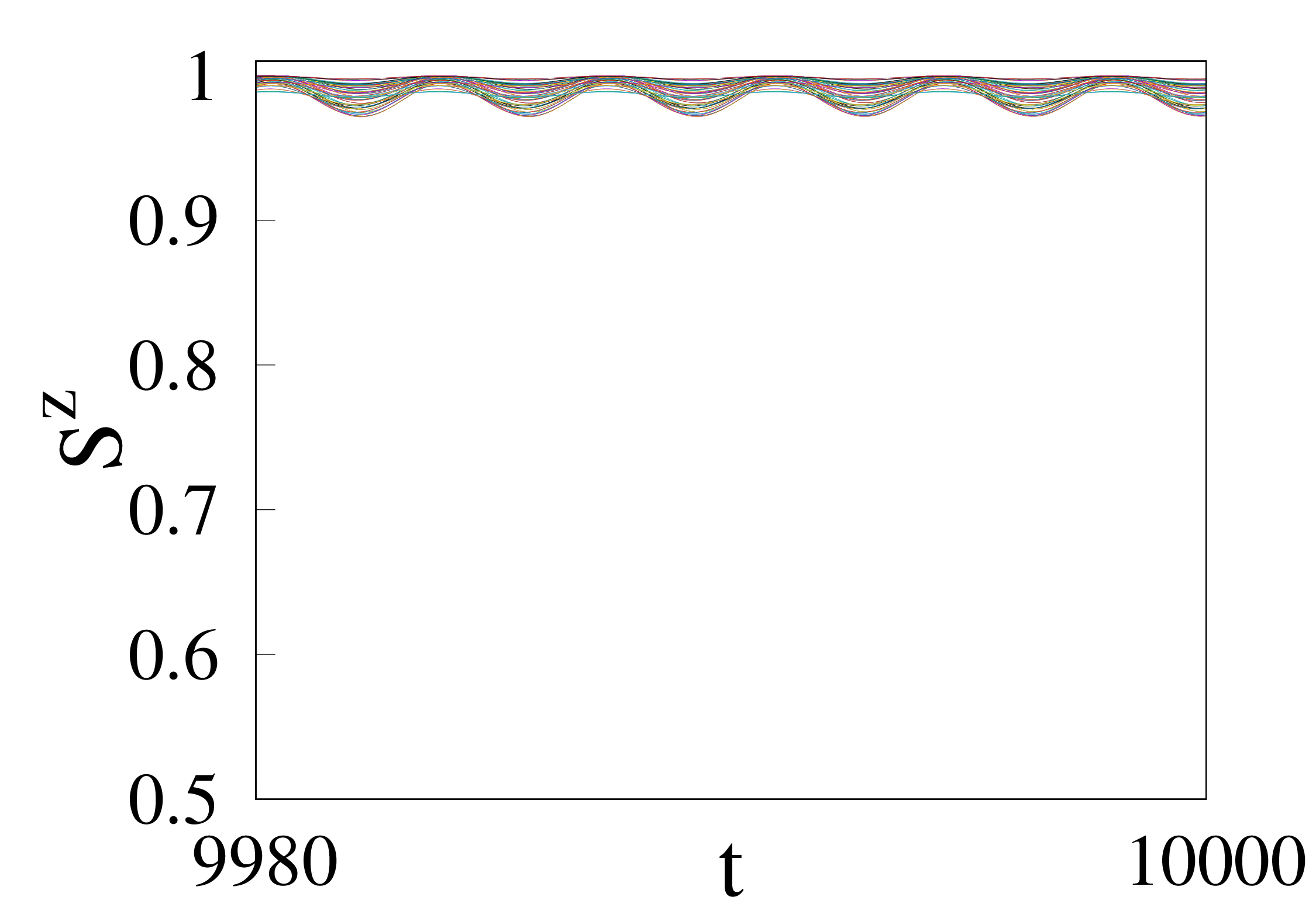}
	\caption{Time evolution of $S^z$ of 100 number of spins for the parameters $A=2,B=2,C=1,Kz=0.1,H_x=0.1,H_y=0,H_z=0,j=0.1,\gamma=1.0,\alpha=0.005,\beta=-0.6$.}
	\label{fig5A}
\end{figure} 

\begin{figure}[h!]
	\centering	\includegraphics[width=1\linewidth]{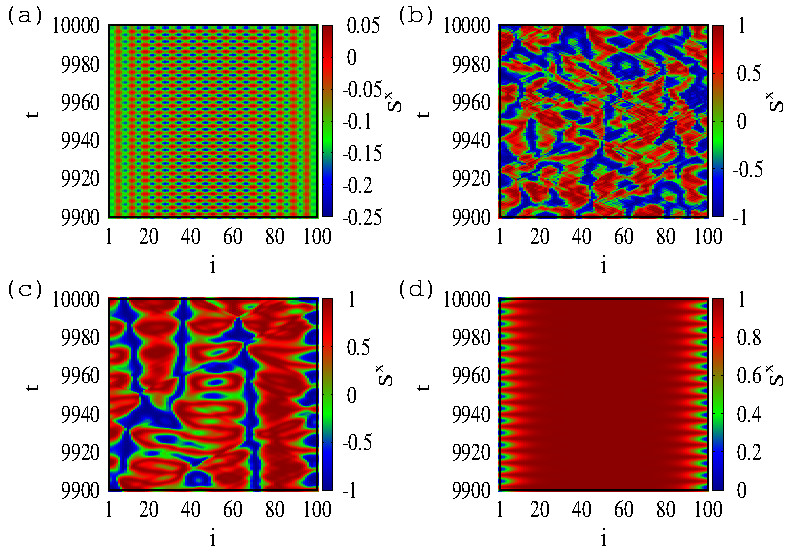}
	\caption{Time evolution of $S^x$ of 100 number of spins for (a) $\alpha$ = 0.005, (b) $\alpha$ = 0.01, (c) $\alpha$ = 0.05, and (d) $\alpha$ = 0.1 corresponding to the parameters $A=2,B=2,C=1,Kz=0.1,H_x=0.1,H_y=0,H_z=0,j=0.1,\gamma=1.0,\beta=-0.6$.}
	\label{Rev_Fig1}
\end{figure}
\begin{figure}[h!]
	\centering	\includegraphics[width=0.9\linewidth]{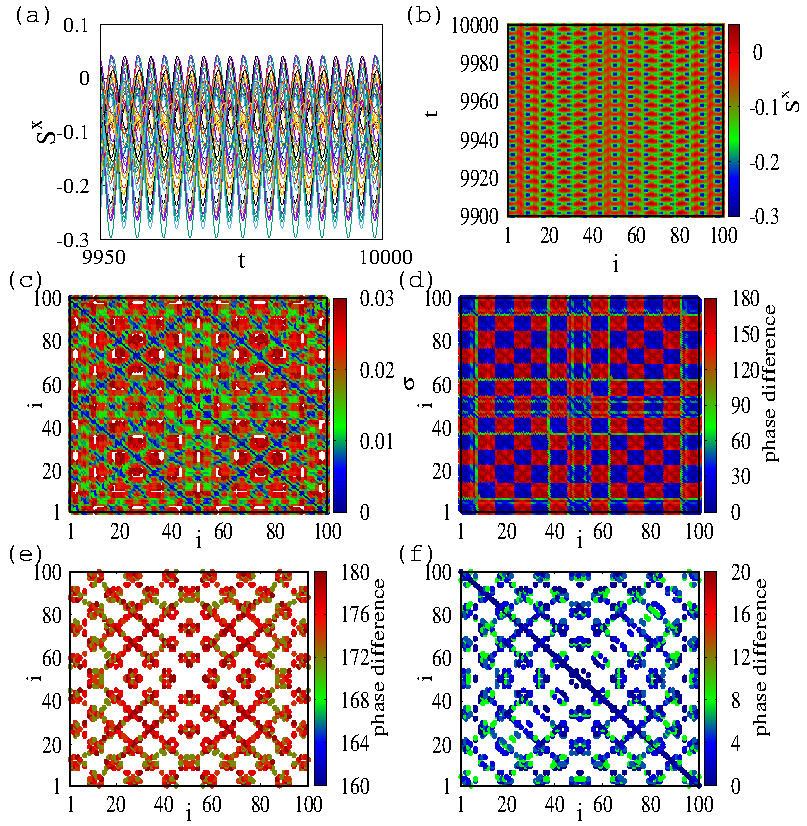}
	\caption{(a,b) Time evolution of the spins, (c) standard deviation and (d-f) phase difference between the different pair of spins. Here, $H_x=0.1,H_y=0, H_z=0.1, j=0.1$, $\beta=-0.6$.}
	\label{fig4}
\end{figure}

From Figs. \ref{fig3}(a) and (b) we can understand that the spins ($i,N+1-i$) are exhibiting complete synchronization. The remaining pairs of spins other than ($i,N+1-i$) oscillate with inphase synchronization.  In Fig. \ref{fig3}(c),  the pair-wise phase difference has been plotted and this exhibits that the spins oscillate with inphase synchronization even though they do not exhibit complete synchronization. The dark blue pattern on the diagonal in Fig. \ref{fig3}(c) implies that the spins around $k=50$ exhibit oscillations close to phase difference $0^\circ$.  Also, each spin exhibits $0^\circ$ phase difference with the neighboring spins.

To validate the numerical results the frequency of the oscillations shown in Fig.\ref{fig2A}, obtained for the set of parameters $A=2,B=2,C=1,K_z=0.1,H_x=0.1,H_y=0,H_z=0,\gamma=1.0,\alpha=0.005,j=0.1,\beta=-0.6$, is analytically calculated. For this purpose, Eq. \eqref{llgs} is written in terms of cartesian coordinates as 
\begin{align}
\frac{dS_k^x}{dt} = &-A\alpha (1-(S_k^x)^2)(S_{k-1}^x+S_{k+1}^x)+B(\alpha S_k^y S_k^x +\gamma S_k^z)(S_{k-1}^y+S_{k+1}^y)\nonumber\\
&+C(\alpha S_k^z S_k^x -\gamma S_k^y)(S_{k-1}^z+S_{k+1}^z) -H_x \alpha (1-(S_k^x)^2)\nonumber\\
&+2 K_z S_k^z (\alpha S_k^x S_k^z -\gamma S_k^y)-j(1-(S_k^x)^2),\label{llg1}\\
\frac{dS_k^y}{dt} = &A(\alpha S_k^x S_k^y -\gamma S_k^z)(S_{k-1}^x+S_{k+1}^x)-B\alpha (1-(S_k^y)^2)(S_{k-1}^y+S_{k+1}^y)\nonumber\\
&+C(\alpha S_k^y S_k^z +\gamma S_k^x)(S_{k-1}^z+S_{k+1}^z) +H_x (\alpha S_k^x S_k^y-\gamma S_k^z)\nonumber\\
&+2 K_z S_k^z (\alpha S_k^y S_k^z +\gamma S_k^x)+j(S_k^x S_k^y+\beta S_k^z),\label{llg2}\\
\frac{dS_k^z}{dt} = &A(\alpha S_k^x S_k^z +\gamma S_k^y)(S_{k-1}^x+S_{k+1}^x)+B(\alpha S_k^y S_k^z -\gamma S_k^x)(S_{k-1}^y+S_{k+1}^y)\nonumber\\
&-C\alpha (1-(S_k^z)^2)(S_{k-1}^z+S_{k+1}^z) +H_x (\alpha S_k^x S_k^z+\gamma S_k^y).\nonumber\\
&-2 K_z\alpha S_k^z (1-(S_k^z)^2)+j(S_k^x S_k^z-\beta S_k^y).\label{llg3}
\end{align}
Using the transformations $S_k^x=\sin\theta_k\cos\phi_k,~S_k^y= \sin\theta_k\sin\phi_k,~S_k^z= \cos\theta_k$, we can derive the dynamical equation in terms of spherical polar coordinates as
\begin{align}
\sin\theta_k \frac{d\phi_k}{dt} &= \cos\phi_k \frac{dS_k^y}{dt} - \sin\phi_k \frac{dS_k^x}{dt}\label{llg4}, \\
\sin\theta_k \frac{d\theta_k}{dt} &= -\frac{dS_k^z}{dt}. \label{llg5}
\end{align}
After substituting Eqs. \eqref{llg1}-\eqref{llg3} in Eqs. \eqref{llg4} and \eqref{llg5} we can obtain
\begin{align}
\sin\theta_k \frac{d\phi_k}{dt} ~=~ &A(\alpha\sin\phi_k-\gamma\cos\theta_k\cos\phi_k)(\cos\phi_{k-1}\sin\theta_{k-1}+\cos\phi_{k+1}\sin\theta_{k+1}) \nonumber\\
&-B(\gamma\cos\theta_k\sin\phi_k+\alpha\cos\phi_k)(\sin\phi_{k-1}\sin\theta_{k-1}+\sin\phi_{k+1}\sin\theta_{k+1})\nonumber\\
&+C\gamma \sin\theta_k(\cos\theta_{k-1}+\cos\theta_{k+1})+H_x(\alpha\sin\phi_k-\gamma\cos\theta_k\cos\phi_k)\nonumber\\
&+K_z\gamma \sin(2\theta_k)+j(\beta \cos\theta_k \cos\phi_k+\sin\phi_k),\label{llg6}\\
\frac{d\theta_k}{dt}~=~&-A(\alpha\cos\theta_k\cos\phi_k+\gamma\sin\phi_k)(\cos\phi_{k-1}\sin\theta_{k-1}+\cos\phi_{k+1}\sin\theta_{k+1})\nonumber\\
&+B(\gamma\cos\phi_k-\alpha\cos\theta_k\sin\phi_k)(\sin\phi_{k-1}\sin\theta_{k-1}+\sin\phi_{k+1}\sin\theta_{k+1})\nonumber\\
&+C\alpha\sin\theta_k(\cos\theta_{k-1}+\cos\theta_{k+1})-H_x(\alpha\cos\theta_k\cos\phi_k+\gamma\sin\phi_k)\nonumber\\
&+K_z\alpha\sin(2\theta_k)-j(\cos\theta_k\cos\phi_k-\beta\sin\phi_k).\label{llg7}
\end{align}
For the inphase synchronization it can be considered that $\phi_{k-1}=\phi_k=\phi_{k+1}$. Also from Fig. \ref{fig5A}, which has been plotted for $S^z(t)$ corresponding to 100 number of spins, we can observe that the value of $S^z(t)$ for all the spins is approximately equal to 1. This indicates that $\theta_k\approx$ constant and $\cos\theta_k\approx 1$, $\sin\theta_k\approx\theta_k$ since $S_k^z=\cos\theta_k$. Hence, $\theta_{k-1}=\theta_k=\theta_{k+1}$. After using the above identities in Eq. \eqref{llg6} we can arrive at
\begin{align}
\theta_k \frac{d\phi_k}{dt} = &\alpha(A-B)\theta_k\sin(2\phi_k)-2\gamma\theta_k(A\cos^2\phi_k+B\sin^2\phi_k-C-K_z)\nonumber\\
&+(H_x \alpha+j)\sin\phi_k+(j\beta-H_x \gamma)\cos\phi_k. \label{llg8}
\end{align}

For the given set of parameters, $A=2,B=2,C=1,K_z=0.1,H_x=0.1,H_y=0,H_z=0,j=0.1,\gamma=1.0,\alpha=0.005,\beta=-0.6$, Eq. \eqref{llg8} can be written as
\begin{align}
\theta_k \frac{d\phi_k}{dt} = -1.8~\theta_k+0.1005\sin\phi_k-0.16\cos\phi_k. \label{llg9}
\end{align}
Integrating Eq. \eqref{llg9} with respect to $\phi_k$ from 0 to $2\pi$, we get,
\begin{align}
	\theta_k \int_0^{2\pi}\frac{d\phi_k}{dt} d\phi_k= -1.8~\theta_k\int_0^{2\pi}d\phi_k+0.1005\int_0^{2\pi}\sin\phi_k d\phi_k-0.16\int_0^{2\pi}\cos\phi_k d\phi_k. \label{llg10}
\end{align}
Here, $\frac{d\phi_k}{dt}=2\pi f$ = constant, where $f$ is the frequency of the oscillations of the spins that exhibit inphase synchronized oscillations. Then the frequency is calculated from Eq. \eqref{llg10} as $|f|= {1.8}/{2\pi}=0.28$, which is exactly equal to the frequency of the oscillations plotted in Fig. \ref{fig2A}, and this validates our numerical results.

The impact of the increase in the value of the damping constant on the dynamics of spins is investigated for the damping constant values $\alpha$ = 0.005, 0.01, 0.05, and 0.1, and their temporal evolution along the $x$-component ($S^x$) have been plotted in Figs.\ref{Rev_Fig1}(a), (b), (c), and (d), respectively. As we can see in Figs.\ref{Rev_Fig1}(a-d), when the damping constant increases above 0.005 the oscillations loose their phase synchronization and exhibit desynchronization. From Figs.\ref{Rev_Fig1}(b-d) we can observe that when the value of damping constant increases the red region increases, which means the number of spins that settles near $S^x$ = 1 enhances. For $\alpha$ = 0.1, all other spins except near both the edges exhibit oscillations as shown in Fig.\ref{Rev_Fig1}(d). Hence it is proved that the increase of damping constant impacts the dynamics of spins in such a way that they loose synchronization in oscillations and reach steady state near $S^x=1$. 
\subsection{Complete, inphase, antiphase and desynchronized oscillations}
In the previous sections we observed that the field-like-torque causes the spins to oscillate in complete and inphase synchronization modes. Here we show that the spins can simultaneously exhibit complete, inphase, antiphase and desynchronization modes as well.  Considering  Figs. \ref{fig4}(a-f), we plot in Figs. \ref{fig4}(a-b) the time evolution of the 100 spins and in (c) the standard deviation between different spins and in (d-f) the phase difference between the different pairs of spins for $\beta = -0.6$. Here, the parameters are set the same as in Fig. \ref{fig3} except that now $H_z$ = 0.1. From the time evolution displayed in Figs. \ref{fig4}(a) and (b) we can observe that the spins oscillate with the character of antiphase synchronization. In addition to the antiphase synchronization the pair of spins ($i,N+1-i$) exhibit complete synchronization. To confirm the complete synchronization between  these pairs, the standard deviation ($\sigma$) is plotted between their values of $S^x(t)$ in Fig. \ref{fig4}(c), where we can see the pair of spins ($i,N+1-i$) represented by the dark blue color exhibiting complete synchronized oscillations. To prove the existence of inphase and antiphase synchronization in the oscillations, the phase difference between all the pairs of spins among the 100 spins is plotted in Fig. \ref{fig4}(d), where the dark blue and red patterns confirm the regions of inphase and antiphase synchronized oscillations, respectively. Fig. \ref{fig4}(e) and (f) show that the spins oscillate with the phase difference of more than 170$^\circ$ and less than 10$^\circ$, respectively, where the white region corresponds to the pairs of spins oscillating desynchronously with phase difference less than 170$^\circ$ and more than 10$^\circ$, respectively.
\begin{figure}[h!]
	\centering	\includegraphics[width=0.6\linewidth]{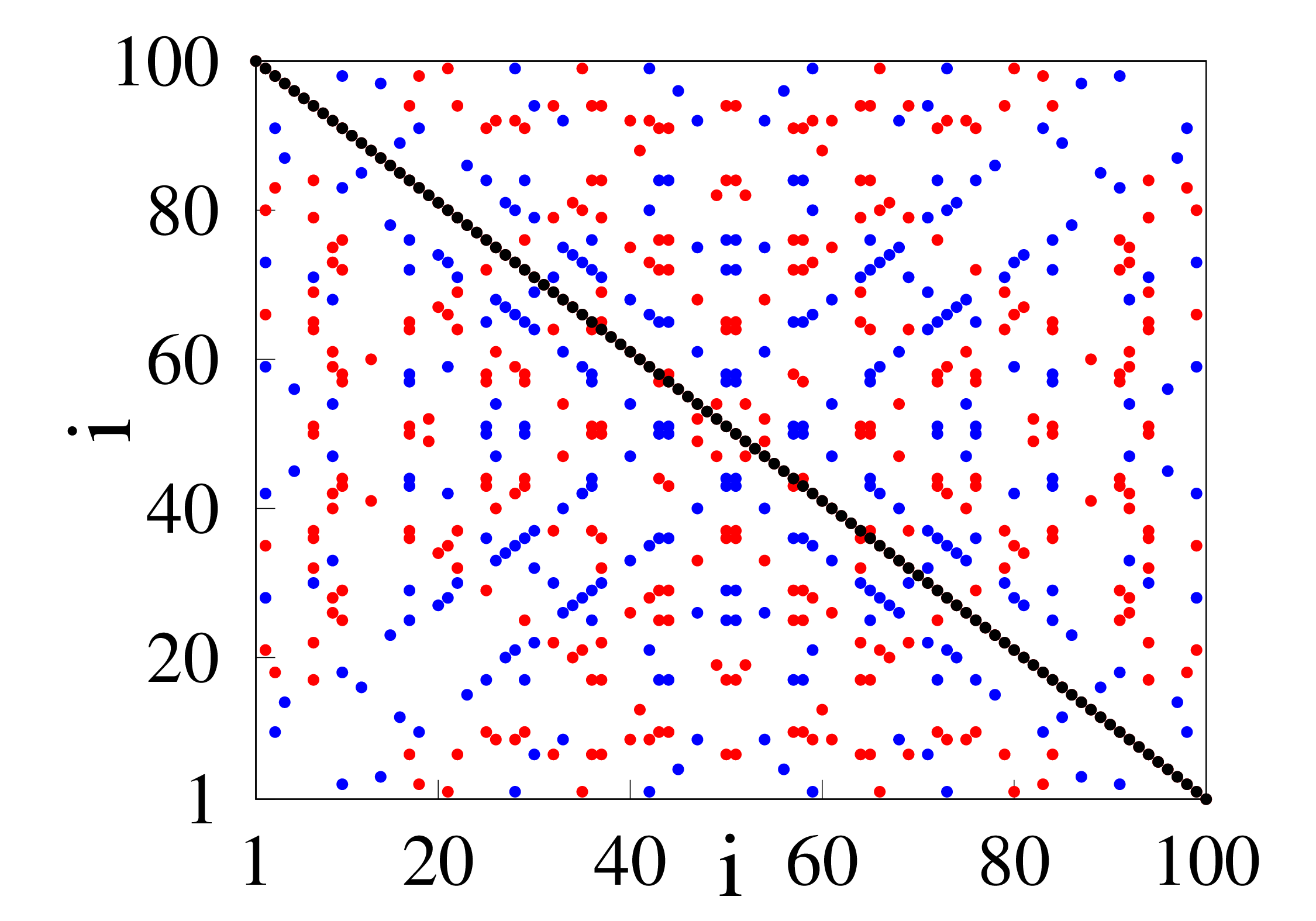}
	\caption{The pair of spins exhibiting inphase ($\sim1^\circ$), antiphase ($\sim180^\circ$), complete synchronized ($\sigma<0.001$) oscillations are plotted with red, blue and black bullets, respectively. Here, $H_x=0.1,H_y=0, H_z=0.1, j=0.1$, $\alpha$=0.005, $\beta=-0.6$.}
	\label{fig4g}
\end{figure} 
From Fig. \ref{fig4} we can conclude that different groups of spins are exhibiting different kinds of oscillations, namely inphase, antiphase, complete and desynchronized oscillations.  To get a clear visualization of the spins between which inphase, antiphase and complete synchronizations exist, in Fig. \ref{fig4g} we have plotted the pairs of spins corresponding to the phase differences $\sim1^\circ$, $\sim180^\circ$, and $\sigma<0.001$, by bullets with colors red, blue and black, respectively. As we observed in Fig. \ref{fig4}, here it is exhibited that the pair of spins ($i,N+1-i$) oscillate in complete synchronization.
\begin{figure}[h!]
\centering	\includegraphics[width=0.9\linewidth]{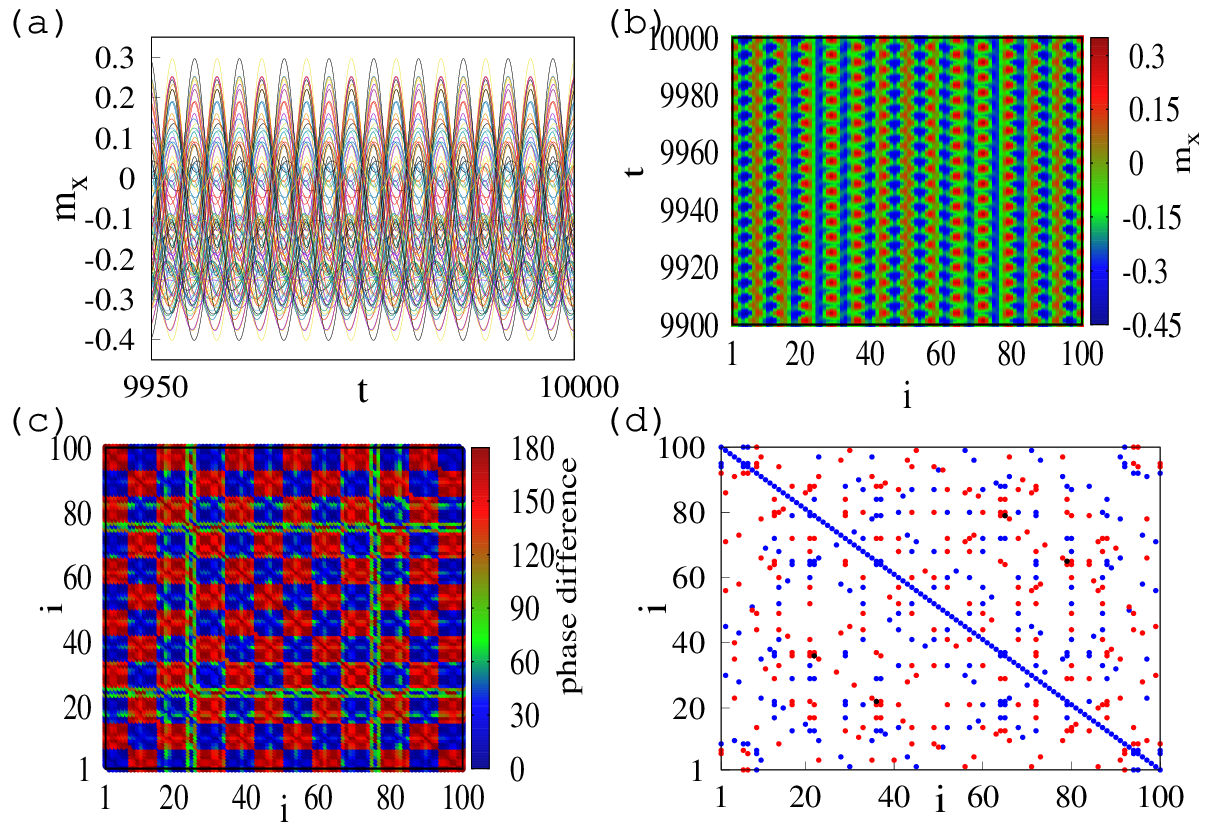}
\caption{(a,b) Time evolution of the spins, (c) Phase difference between the different pairs of spins, and (d) inphase (red), antiphase (blue), and complete synchronization (black). Here, $H_x=0.1,H_y=0.2, H_z=0.3, j=0.1$, $\beta=-0.6$.}
\label{fig6}
\end{figure} 

Further, we can show the possibility of antiphase synchronized oscillations between the pairs ($i,N+1-i$) instead of inphase synchronization for the values of fields $H_x=0.1,H_y=0.2,H_z=0.3$. The time evolution is plotted in Figs. \ref{fig6}(a) and (b), where the antiphase nature is clearly visible. In Fig. \ref{fig6}(c) the phase difference is plotted for different pairs of spins  with different color shades, which confirms the existence of inphase, antiphase and desynchronized oscillations. Also, the spins exhibiting inphase, antiphase and complete synchronization are plotted with red, blue and black bullets, respectively in Fig. \ref{fig6}(d). 

\section{Conclusions}
We have numerically simulated the dynamics of spins in a one-dimensional anisotropic Heisenberg ferromagnetic spin chain for their different modes of self-oscillations by solving Landau-Lifshitz-Gilbert-Slonczewski equation numerically. The dynamical equation for the `$k$'-th spin has been written from the Hamiltonian corresponding to the $N$ number of spins. The self-oscillations in spins are achieved for the specific values of external field and current.  It is shown that when the number of spins becomes large the synchronization is lost between them and they oscillate desynchronously. However, the synchronization can be restored with the inclusion of field-like torque and we have proved the simultaneous existence of inphase and complete synchronization in the oscillations. The complete synchronization is shown between the pair of spins ($i,N+1-i$), which has been confirmed by plotting standard deviation between their  oscillations. Additionally, we have confirmed the coexistence of inphase, antiphase, complete and desynchronized modes of oscillations among the spins when the magnetic field is applied in both directions $x$ and $z$. Further, we have shown that the antiphase synchronization between the pairs  ($i,N+1-i$) occurs instead of complete synchronization by applying the field in all the directions $x$, $y$, and $z$. The frequency of the inphase synchronized oscillations is analytically derived and it matches with the numerical values.

{\flushleft \bf ACKNOWLEDGMENTS}

M.L. wishes to acknowledge the ANRF award of a ANRF-SERB National Science Chair under Grant No. NSC/2020/00029. R.A. would like to thank SRM TRP Engineering College, India, for their financial support, vide number SRM/TRP/RI/005.  Work of A.S. was supported by the U.S. Department of Energy.

{\flushleft \bf APPENDIX: Comparison of numerical parameters with realistic material parameters}

Here, we briefly outline the procedure for deriving the expressions for the current density $j$ and the field-like torque coefficient $\beta$ by comparing Eq.\eqref{llgs} with the standard form of the Landau–Lifshitz–Gilbert–Slonczewski (LLGS) equation for a spin-torque nano-oscillator (STNO), which comprises a ferromagnetic free layer and a pinned layer separated by a nonmagnetic conducting spacer.

{
As discussed in Section II, the LLGS equation for a spin in the presence of field, current and field-like torque with the simplification that all the spins orient in the same direction is given by is given by
\begin{align}
\frac{d\bf S}{dt} = -\gamma ~{\bf S}\times{\bf H}_{eff} - \alpha~ {\bf S}\times({\bf S}\times{\bf H}_{eff})+j~ {\bf S}\times({\bf S}\times {\bf e}_x)+j\beta ~{\bf S}\times {\bf e}_x,
\tag{A.1}\label{A1}
\end{align}
where the effective field is given by
\begin{align}
	{\bf H}_{eff} = (2A S^x+H_x){\bf e}_x + (2B S^y+H_y){\bf e}_y + (2S^z(C+K_z)+H_z){\bf e}_z.
	\tag{A.2}\label{A2}
\end{align}
One can deduce the following equation from Eq.\eqref{A1} using the orthogonality condition ${\bf S}\cdot\frac{d{\bf S}}{dt}$ = 0 as
\begin{align}
\frac{\alpha}{\gamma} ~{\bf S}\times \frac{d\bf S}{dt} = -\alpha~ {\bf S}\times({\bf S}\times{\bf H}_{eff})+\frac{\alpha^2}{\gamma} ~{\bf S}\times{\bf H}_{eff}-\frac{\alpha j}{\gamma}~ {\bf S}\times {\bf e}_x + \frac{\alpha j \beta}{\gamma} ~{\bf S}\times ({\bf S}\times {\bf e}_x). \tag{A.3} \label{A3}
\end{align}
By substituting $-\alpha {\bf S}\times({\bf S}\times{\bf H}_{eff})$ from Eq.\eqref{A3} in Eq.\eqref{A1} we can arrive at
\begin{align}
\frac{\gamma^2}{\gamma^2+\alpha^2}~\frac{d\bf S}{dt} = -\gamma ~ {\bf S}\times{\bf H}_{eff} + \frac{\gamma\alpha}{\gamma^2+\alpha^2}~{\bf S}\times \frac{d\bf S}{dt} +\gamma j\frac{\gamma-\alpha\beta}{\gamma^2+\alpha^2}~{\bf S}\times ({\bf S}\times {\bf e}_x)+\gamma j\frac{\gamma\beta+\alpha}{\gamma^2+\alpha^2}~{\bf S}\times {\bf e}_x \tag{A.4}\label{A4}
\end{align}
With a rescaling of time $t \rightarrow \frac{\gamma^2}{\gamma^2+\alpha^2}t$, we get
\begin{align}
\frac{d\bf S}{dt} = -\gamma ~{\bf S}\times {\bf H}_{eff} + \frac{\alpha}{\gamma}~{\bf S}\times \frac{d\bf S}{dt} +\gamma j\frac{\gamma-\alpha\beta}{\gamma^2+\alpha^2} {\bf S}\times ({\bf S}\times {\bf e}_x) +\gamma j\frac{\gamma\beta+\alpha}{\gamma^2+\alpha^2} {\bf S}\times {\bf S}_p \tag{A.5}\label{A5}.
\end{align}
The standard form of LLGS equation used to investigate the dynamics of the free layer's magnetization ${\bf m}$ in an STNO is given by \cite{tani,arun}
\begin{align}
\frac{d\bf m}{dt} = -\gamma' {\bf m}\times {\bf H}_{eff}' +\alpha' ~{\bf m}\times \frac{d\bf m}{dt}+ \gamma' H_s~{\bf m}\times ({\bf m} \times {\bf e}_x) +\gamma' H_s \beta'~{\bf m}\times {\bf e}_x \tag{A.6}\label{A6},
\end{align} 
where
\begin{align}
{\bf H}_{eff}' = (K_x' m_x + H_x'){\bf e}_x + (K_y' m_y + H_y') {\bf e}_y + (K_z' m_z + H_z') {\bf e}_z \tag{A.7}\label{A7}
\end{align}
and 
\begin{align}
	H_S = \frac{\hbar \eta I}{2 e M_s V} \tag{A.8}\label{A8}
\end{align}
Here, ${\bf H}_{eff}'$ includes fields due to external fields $H_x'$, $H_y'$, $H_z'$ along $x$, $y$, $z$ directions, respectively, and anisotropy fields $K_x'$, $K_y'$, $K_z'$ along $x$, $y$, $z$ directions, respectively. $\gamma'$ is the gyromagnetic ratio, $\alpha'$ is the damping constant, $\beta'$ is the field-like torque related with STNO, $\hbar(= h/2\pi)$ is the reduced Planck’s constant, $\eta$ is the dimensionless parameter which determines the magnitude of the spin transfer torque, $I$ is the current flowing through the free layer, $e$ is the charge of the electron, $M_s$ is the saturation magnetization and $V$ is the volume of the free layer.}

{
By comparing Eqs.\eqref{A5} and \eqref{A6} we can obtain
\begin{align}
\alpha' = \alpha/\gamma,\hskip 0.5cm H_s = \frac{\hbar\eta I}{2eM_s V} = \frac{j(\gamma-\alpha\beta)}{\gamma^2+\alpha^2},\hskip0.5cm H_s\beta' = \frac{j(\gamma\beta+\alpha)}{\gamma^2+\alpha^2} \tag{A.9}\label{A9}
\end{align}
and
\begin{align}
K_x' = 2A,\hskip0.2cm K_y' = 2B, \hskip0.2cm K_z' = 2(C+K_z),\hskip0.2cm H_x'=H_x, \hskip0.2cm H_y'=H_y, \hskip0.2cm H_z'=H_z,\hskip0.2cm \gamma'=\gamma \tag{A.10}\label{A10}
\end{align}
From Eq.\eqref{A9} we can derive the current through the STNO as
\begin{align}
I = \frac{2 e M_s V j}{\hbar \eta}\left(\frac{\gamma-\alpha\beta}{\gamma^2+\alpha^2}\right) \tag{A.11}\label{A11}
\end{align}
and field-like torque corresponding to STNO is
\begin{align}
	\beta' = \frac{\gamma\beta+\alpha}{\gamma-\alpha\beta}\tag{A.12}\label{A12}.
\end{align}
}

{\flushleft \bf  CRediT authorship contribution statement}

{\bf R. Arun:} Conceived the idea and did the analytical work, carried out the numerical work, Discussed, interpreted the results, Wrote the manuscript. {\bf M. Lakshmanan:} Conceived the idea and did the analytical work, Discussed, interpreted the results, Wrote the manuscript.  {\bf Avadh Saxena:} Conceived the idea and did the analytical work, Discussed, interpreted the results, Wrote the manuscript

{\flushleft \bf Author Declarations}

The authors declare that they have no known competing financial interests or personal relationships that could have appeared to influence the work reported in this paper.

{\flushleft \bf Conflict of Interest}

The authors have no conflicts to disclose.

{\flushleft \bf Data Availability}

The data that support the findings of this study are available from the corresponding author upon reasonable request.

%\nocite{*}
%\bibliography{manuscript}

%aipnum4-2.bst 2019-01-14 (MD) hand-edited version of apsrev4-1.bst
\providecommand{\noopsort}[1]{}\providecommand{\singleletter}[1]{#1}%
\end{document}